\documentclass[letterpaper, 10 pt, conference]{IEEEtran}  

\IEEEoverridecommandlockouts                              

\usepackage[utf8]{inputenc}
\usepackage[T1]{fontenc}
\usepackage{graphicx}

\title{\LARGE \bf
The Causality Inference of Public Interest in Restaurants and Bars on COVID-19 Daily Cases in the US: A Google Trends Analysis
}

\author{Milad Asgari Mehrabadi$^{1}$, Nikil Dutt$^{1,2}$, and Amir M. Rahmani$^{2,3}$
\thanks{$^{1}$Department of Electrical Engineering and Computer Science, University of California Irvine.
        {\tt\small masgarim@uci.edu}}
\thanks{$^{2}$Department of Computer Science, University of California Irvine.
        {\tt\small}}%
\thanks{$^{3}$School of Nursing, University of California Irvine.
        {\tt\small}}%
}

\begin{document}

\maketitle
\thispagestyle{empty}
\pagestyle{empty}

\begin{abstract}

The COVID-19 coronavirus pandemic has affected virtually every region of the globe. At the time of conducting this study, the number of daily cases in the United States is more than any other country, and the trend is increasing in most of its states. Google trends provide public interest in various topics during different periods. Analyzing these trends using data mining methods might provide useful insights and observations regarding the COVID-19 outbreak. The objective of this study was to consider the predictive ability of different search terms (i.e., bars and restaurants) with regards to the increase of daily cases in the US. In particular, we were concerned with searches for dine-in restaurants and bars. Data were obtained from Google trends API and COVID tracking project. We considered the causation of two different search query trends, namely restaurant and bars, on daily positive cases in top-10 states/territories of the United States with the highest and lowest daily new positive cases. In addition, to measure the linear relation of different trends, we used Pearson correlation. Our results showed for states/territories with higher numbers of daily cases, the historical trends in search queries related to bars and restaurants, which mainly happened after re-opening, significantly affect the daily new cases, on average. California, for example, had most searches for restaurants on June 7th, 2020, which affected the number of new cases within two weeks after the peak with the P-value of .004 for Granger’s causality test. Although a limited number of search queries were considered, Google search trends for restaurants and bars showed a significant effect on daily new cases for regions with higher numbers of daily new cases in the United States. We showed that such influential search trends could be used as additional information for prediction tasks in new cases of each region. This prediction can help healthcare leaders manage and control the impact of COVID-19 outbreaks on society and be prepared for the outcomes.

\end{abstract}


\section{INTRODUCTION}
The entire world has been affected significantly by a global virus pandemic. The first case of this virus was reported in China during December 2019, and the first case outside China was discovered in January 2020 \cite{timeline}. During February, the World Health Organization called this virus COVID-19 \cite{guo2020origin}.

Worldwide, there have been 14.4M confirmed cases, with 604K deaths, as of 19th of July 2020 \cite{worldometers}. The United States of America, with 3.83M confirmed cases and 143k deaths, is the most affected country around the world. In some states, the numbers are still increasing (e.g., California), while in some other states such as New York, the peak has passed, and the average daily new cases are decreasing. 

Due to the rapid spreading of this virus, finding effective reasons can play a significant role in prevention policies. Using  data mining and time series analysis methods, it is possible to investigate the impact of different phenomena on time series data. In economics, as an example, there are different studies that model the temporal relationship of two or more time series (e.g., the relationship between oil and gold price) using the same methods \cite{vsimakova2011analysis}. 

Google search trends can be useful for reflecting public interests/concerns during different periods \cite{ayyoubzadeh2020predicting}. During the COVID-19 outbreak, different studies have investigated the correlation of web-based data and cases of this virus. Kutlu et al. \cite{kutlu2020analysis} investigated the correlation of dermatological diseases obtained by specific Google search trends with the COVID-19 outbreak. In addition, Google trends have been utilized to predict and monitor COVID-19 cases around the world \cite{ayyoubzadeh2020predicting, li2020retrospective, effenberger2020association, ciaffi2020google, mavragani2020tracking, husnayain2020applications}. Multiple studies analyzed the data related to the US to correlate the search trends and COVID-19 cases \cite{yuan2020trends, hong2020population, walker2020use, husain2020covid, jacobson2020flattening, rajan2020association}. However, these studies did not consider the predictive ability of search trends on future confirmed cases.

In this paper, we considered the causality effect and predictive ability of search terms related to bars and restaurants on the daily new cases of the US in different regions. Along with the linear correlation analysis between search trends and COVID-19 cases, we have utilized the statistical causality methods to investigate the influential confidence of these methods on COVID-19 daily new cases.
 
\section{Methods}

\subsection{Datasets}
For our analysis, we obtained the daily cases of COVID-19 in the US using the COVID tracker project \cite{covidtracking}. This project compiles the daily statistics, including the number of positive/negative tests, hospitalization, available ventilators, and the number of deaths from each US state and territory. For this study, we considered the data of approximately three months starting April 09, 2020, to July 07, 2020, which contains 5040 samples for 56 states/territories. 

We used Google Trends to obtain the public interest in bars and restaurant categories with daily resolution. We used the most popular query for each category from April 09, 2020, to July 07, 2020, for 45 available regions in Google trends API. For restaurants and bars, we chose “dine-in restaurants that are open near me” and “bars near me”, respectively. Google trend does not provide the number of queries per day. Instead, it provides a normalized number between 0 and 100, where 0 refers to “low volume of data for the query” while 100 refers to the “highest popularity for the term” \cite{googletrends}. To be consistent with Google trend values, we normalized the US daily new cases between 0 and 100 in our analysis.

Aggregating data from Google trends results and COVID-19 daily cases, and removing missing values, resulted in available data for 45 regions in the US. We categorized our analysis to two different groups: First,  top-10 states/territories with the highest number of daily new cases as of July 7th, 2020 which consist of Texas (TX), Florida (FL), California (CA), Arizona (AZ), Georgia (GA), Louisiana (LA), Tennessee (TN), North Carolina (NC), Washington (WA) and Pennsylvania  (PA). Second, top-10 states/territories with the lowest number of daily new cases as of July 7th, 2020: Kansas (KS), Hawaii (HI), New Hampshire (NH), Maine (ME), West Virginia (WV), Rhode Island (RI), Connecticut (CT), Montana (MT), Nebraska (NE) and Delaware (DE).

\subsection{Correlation and Causation}
To analyze the linear correlation of two time-series, the Pearson correlation has been utilized. The value of such a correlation ranges from -1 to 1, which shows a negative and positive correlation, respectively. Our analysis measured the Pearson correlation between the trends of search queries (i.e., restaurants and bars) and the daily new cases of COVID-19 in each state.

In addition, we used Granger’s causality \cite{granger1969investigating} to model the influence of a time series’ past values on the new values of another time series. Granger’s causality tests whether the past values of a time series X cause the current values of another time series Y. Hence, in this study, the null hypothesis is that X’s past values do not affect Y’s current values. If the P-value is less than the marginal value (.05), we can reject the null hypothesis. In our analysis, we reported P-values for each aforementioned search query’s influence on the daily new cases. One of the main assumptions of modeling the influence of time series on each other is their stationarity. To test such a characteristic, we used the Augmented Dickey-Fuller (ADF) test \cite{fuller2009introduction} as our unit root test. This test determines the effect of a trend in the creation of the time series. In other words, it determines how strongly a trend defines a time series. The alternative hypothesis in the ADF test is the stationarity of the time series. 

In this study, since the time series were not stationary, we applied first differencing on search trends and second differencing on daily new cases to make all of the three series stationary. For statistical analysis, we used the Python Statsmodel package \cite{seabold2010statsmodels}.

\subsection{Vector Autoregression}
In our study, we leveraged the fact that search trends might impact the daily new cases in the future; hence a Vector Autoregression (VAR) \cite{johansen1995likelihood} model for each region was fitted to the data. A VAR model takes into account the influence of the past values of time series X and Y on current values of time series Y with a given lag order. Lag order with the lowest Akaike's Information Criterion (AIC) was picked in this study. Since symptoms may appear within 2-14 days after exposure to the COVID-19 virus \cite{symptoms}, a maximum of 14 lags was used. The equation for the VAR model with two lags is summarized below:
\[
Y_t = \alpha + \beta_1X_{t-1} + \beta_2X_{t-2} + \beta_1'Y_{t-1} + \beta_2'Y_{t-2} + \epsilon_t
\]
In this model, $Y_t$ represents the value of time series $Y$ at time $t$, which consists of a combination of previous lag values from $Y$ and $X$ with different weights $\beta,\beta'$ and random white noise, $\epsilon_t$. We fitted a VAR model to perform Granger’s causality test.

\subsection{Long Short-Term Memory}
Long Short-Term Memory (LSTM) \cite{hochreiter1997long} models are a type of recurrent neural network useful for time series prediction. These models capture the long term effect of time series as well as their most recent values. In this study, we utilized LSTMs to predict the daily new cases using two sets of features: 1) the historical values of the new cases time series and 2) using additional information from searching query time series. We used 70\% of the data for training, and the rest were used for evaluation of the model. Root mean square error (RMSE) was selected as the performance metric. RMSE can be calculated as follows:
\[
RMSE = \sqrt{\frac{1}{N}\sum(\hat{Y} - Y)^2}
\]
In this equation, $N$ is the number of samples, $\hat{Y}$ is the predicted value, and $Y$ is the actual value of the time series. 

\begin{figure}[!t]
\centering
\includegraphics[width=0.5\textwidth]{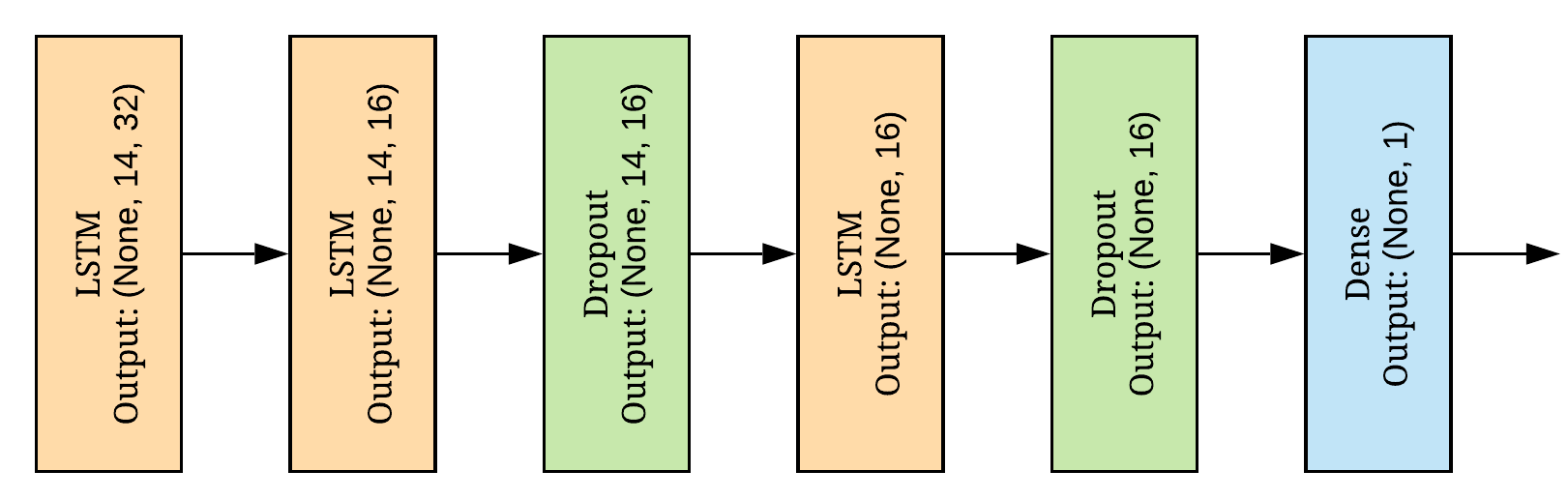}
\caption{The proposed model architecture.}
\label{fig:arch}
\end{figure}

The architecture of the used model is illustrated in Fig. \ref{fig:arch}. It consists of 3 LSTM layers along with dropout layers, and a fully connected layer at the end. Dropout layers were utilized to avoid overfitting, which is a typical problem in Machine Learning tasks. To train such a model, we used the TensorFlow package of Python.

\section{Results}
\begin{figure*}[!t]
\centering
\includegraphics[width=\textwidth]{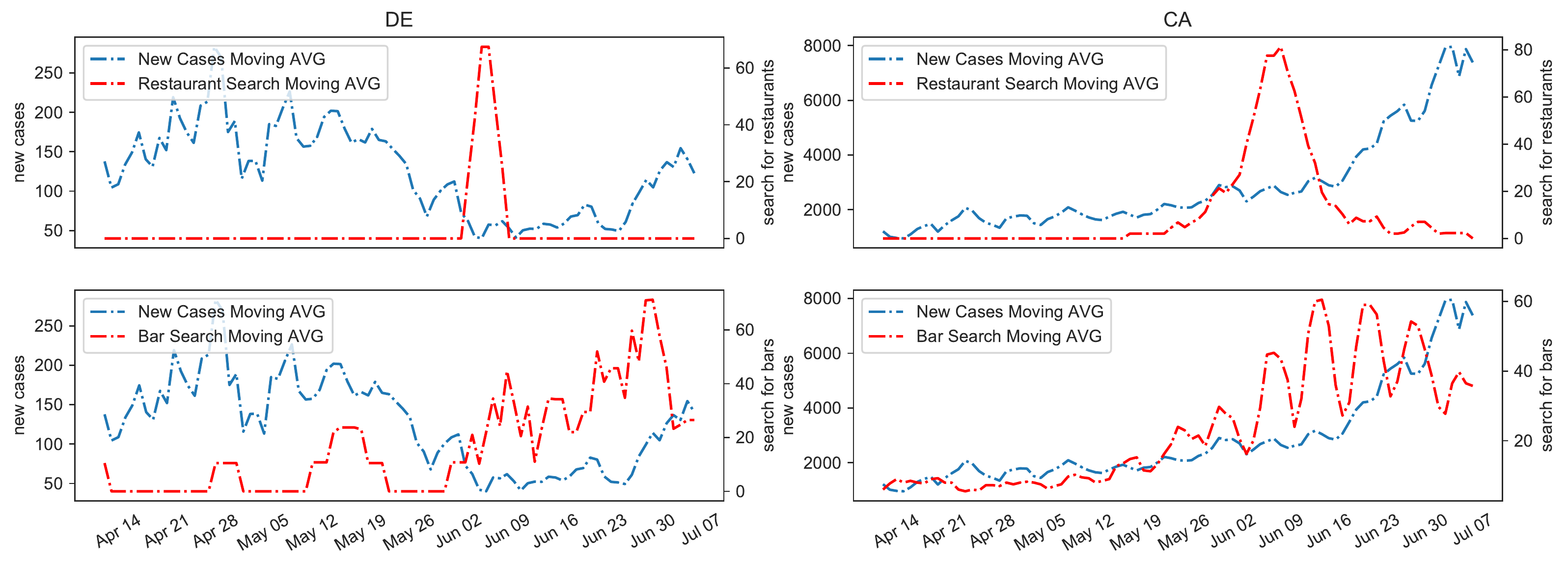}
\caption{Comparison between the effect of California (CA) and Delaware (DE) restaurant and bar search trends on daily cases.}
\label{fig:comparison}
\end{figure*}

{\em Observation.} Search trends and the daily new cases can be different in each state/territory. Hence, the significance level of influence of search-related time series on the current values of daily new cases is different in each region. For the sake of comparison, Fig. \ref{fig:comparison} illustrates the moving average trend of “bar and restaurant” searches as well as the daily new cases in California (CA) and Delaware (DE).

Fig. \ref{fig:comparison} shows for regions such as CA, there was a steep rise of restaurant searches peaking on June 7th. The daily new cases have a drastic increase within two weeks of such a peak. Considering the bar searches in CA, the plot shows an increasing trend with peak value on June 13th.  However, in DE, the daily new cases are not profoundly affected by such search trends. One reason could be the lower population as it is reflected in the number of daily new cases. The other reason can be the high number of new daily cases in California at the time of re-opening restaurants and bars (+2000). 

{\em Granger’s Causality Test.} Due to the inclusion of a large number of states/territories in our analysis (45), we picked top-10 regions in the US with the highest and lowest daily new cases as of July 7th, 2020. Table \ref{tbl:high-cause} summarizes the P-values for testing the null hypothesis (the coefficients corresponding to past values of the second time series are zero) for the first group. P-values below .05 represents the rejection of the null hypothesis, which shows the effect of searching queries on daily new cases for each region.

\begin{table*}[!t]
\centering
\caption{Granger’s causality test (P-values) on daily new cases for top-10 regions with the most daily new cases in the US.}
\label{tbl:high-cause}
\resizebox{\textwidth}{!}{%
\begin{tabular}{|c|c|c|c|c|c|c|c|c|c|c|}
\hline
causing -> caused & Texas & Florida & California & Arizona & Georgia & Louisiana & Tennessee & North Carolina & Washington & Pennsylvania\\
\hline
Restaurant search -> new cases & 0.108 &0.35 &0.004 &0.003 &0.30 &<0.001 &0.091 &0.53 &<0.001 &0.108\\
Bar search -> new cases & 0.019 &0.15 & <0.001 &0.042 &0.001 &<0.001 &0.075 &0.19 &0.016 &0.013\\
\hline
\end{tabular}
}
\end{table*}

\begin{figure}[!t]
\centering
\includegraphics[width=0.5\textwidth]{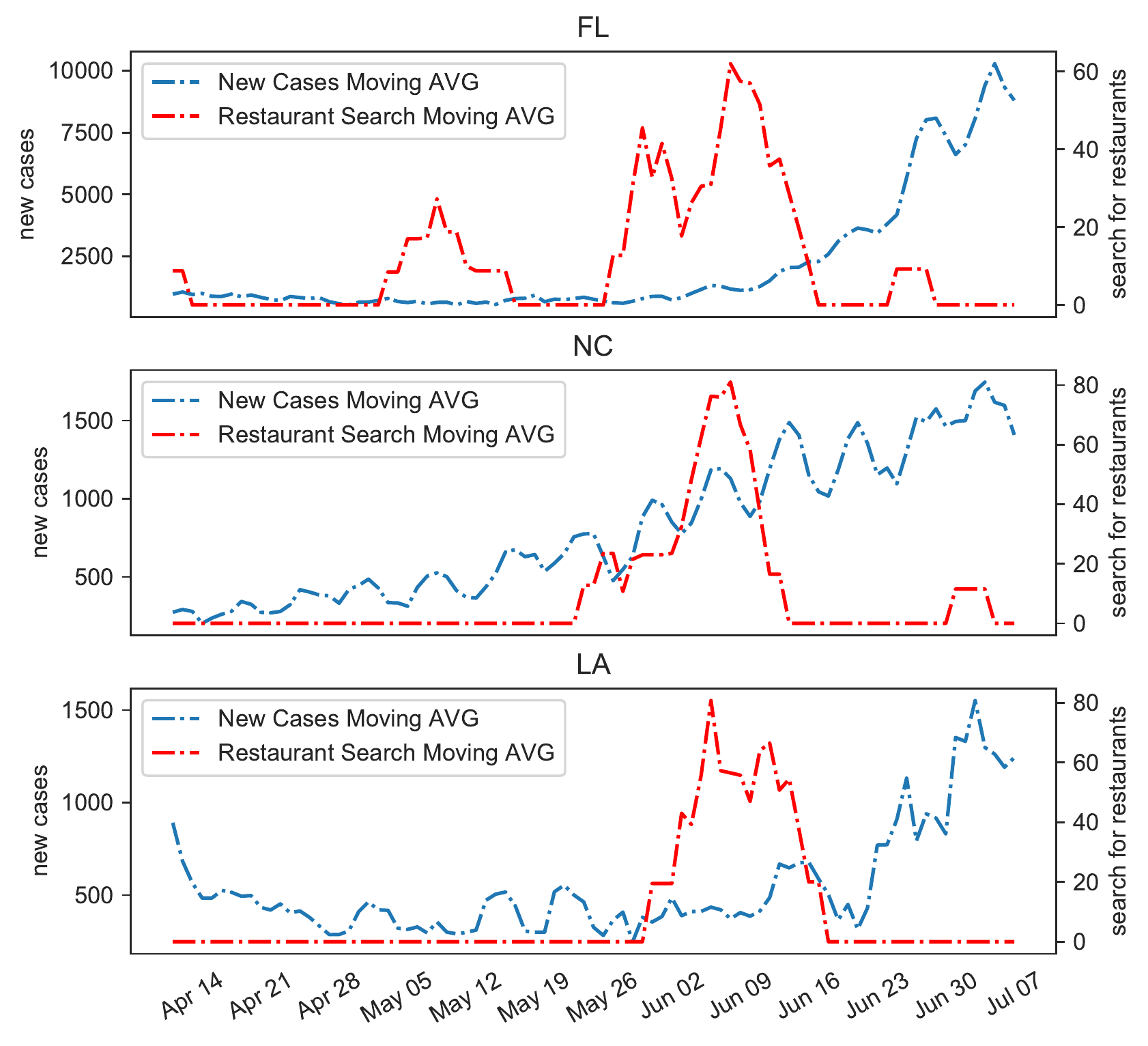}
\caption{Comparison of restaurant search effect on daily new cases in Florida (FL), North Carolina (NC), and Louisiana (LA).}
\label{fig:comparison2}
\end{figure}

Based on Table \ref{tbl:high-cause}, California has small P-values, which shows the influence of the aforementioned search queries on daily new cases; hence, they can be used to predict daily new cases. Florida and North Carolina are two examples of states that the effect of restaurants is rejected with the Granger’s Causality Test; however, Louisiana clearly is affected by restaurant searches. Fig. \ref{fig:comparison2} illustrates the moving average of daily new cases and restaurant search trends for these three states. Based on Fig. \ref{fig:comparison2}, the high P-value for Florida is because of the first peak in the restaurant search, which did not change the daily new cases trends. North Carolina has an overall increasing trend, causing the effect of the search to be marginal. However, Louisiana is influenced by the sudden changes in restaurant search trends, which clearly have affected the daily new cases.

Similarly, Table \ref{tbl:low-cause} summarizes the P-values for Granger’s causality test for the second group (i.e., top-10 regions with the lowest daily new cases).

\begin{table*}[!t]
\centering
\caption{Granger’s causality test (P-values) on daily new cases for top-10 regions with the lowest daily new cases in the US.}
\label{tbl:low-cause}
\resizebox{\textwidth}{!}{%
\begin{tabular}{|c|c|c|c|c|c|c|c|c|c|c|}
\hline
causing -> caused & Kansas & Hawaii & New Hampshire & Maine & West Virginia & Rhode Island & Connecticut & Montana & Nebraska & Delaware\\
\hline
Restaurant search -> new cases & 0.99 & <0.001 &0.88 &0.077 &0.081 &0.54 & 0.99 &<0.001 &0.99 &1.0\\
Bar search -> new cases & 0.014 &0.001 & 0.05 &0.11 &0.45 &0.28 & 0.008 &0.073 &0.083 &<0.001\\
\hline
\end{tabular}
}
\end{table*}

Comparison between Tables \ref{tbl:high-cause} and \ref{tbl:low-cause} shows that regions with higher daily cases are more affected by restaurant and bar searches on average.

{\em Pearson Correlation}
To show the linear relationship of time series, the Pearson correlation is utilized. Tables \ref{tbl:high-corr} and \ref{tbl:low-corr} summarize these correlations with corresponding P-values for each group. Based on these two tables, the linear correlation between the search trends related to bars/restaurants and daily new cases in regions with a higher number of daily cases is more substantial, on average, compared to regions with lower daily cases.

\begin{table*}[!t]
\centering
\caption{Pearson correlation between search trends and daily new cases for top-10 regions with the most daily new cases in the US.}
\label{tbl:high-corr}
\resizebox{\textwidth}{!}{%
\begin{tabular}{|c|c|c|c|c|c|c|c|c|c|c|}
\hline
Correlation (r [P-value]) & Texas & Florida & California & Arizona & Georgia & Louisiana & Tennessee & North Carolina & Washington & Pennsylvania\\
\hline
Restaurant vs. New cases & -0.17 [0.111] &-0.19 [0.072] &-0.0 [0.966] &-0.11 [0.301] &-0.2 [0.065] &-0.13 [0.235] &-0.18 [0.081] &0.17 [0.107] &-0.11 [0.29] &-0.23 [0.027]  \\
Bar vs. New cases & 0.11 [0.289] &0.41 [<0.001] &0.47 [0.0] &0.31 [0.003] &0.31 [0.003] &0.12 [0.264] &0.39 [<0.001] &0.73 [<0.001] &0.13 [0.209] &-0.52 [<0.001]  \\
\hline
\end{tabular}
}
\end{table*}

\begin{table*}[!t]
\centering
\caption{Pearson correlation between search trends and daily new cases for top-10 regions with the lowest daily new cases in the US.}
\label{tbl:low-corr}
\resizebox{\textwidth}{!}{%
\begin{tabular}{|c|c|c|c|c|c|c|c|c|c|c|}
\hline
Correlation (r [P-value]) & Kansas & Hawaii & New Hampshire & Maine & West Virginia & Rhode Island & Connecticut & Montana & Nebraska & Delaware\\
\hline
Restaurant vs. New cases & -0.05 [.62] & -0.08 [.43] & -0.08 [.45]&	-0.08 [.42] &	0.09 [.35]&	-0.08 [.42]& -0.06 [.55]&-0.01 [.85]&	-0.05 [.61]&	-0.17 [.097]  \\
Bar vs. New cases & -0.20 [.057]&	0.22 [.030]&	-0.11 [.27]	&0.13 
[.21]&	0.11 [.28]&	-0.61 [<.001]&	-0.22 [.035]&	0.19 [.070]	&0.007 [.94]&	-0.18 [.087]
  \\
\hline
\end{tabular}
}
\end{table*}

{\em New Cases Prediction.} We used LSTM models to predict the value of daily new cases for a given region. We utilized the search trend time series as additional information to adjust the predicted values. The RMSE scores for test data for top-10 highest and lowest daily new cases are summarized in Tables \ref{tbl:high-eval} and \ref{tbl:low-eval}. These tables show the results for: 
\begin{enumerate}
    \item the baseline model which uses only the past values of new cases time series for the prediction, 
    \item the model that uses the past values of restaurant searches along with the past values of new cases time series, 
    \item and, finally, the model that combines the information from the daily cases and bar searches time series.
\end{enumerate}

\begin{table*}[!t]
\centering
\caption{RMSE scores for new cases time series (Baseline), Baseline + Restaurants time series, and Baseline + Bars time series for top-10 regions with the most daily new cases in the US.}
\label{tbl:high-eval}
\resizebox{\textwidth}{!}{%
\begin{tabular}{|c|c|c|c|c|c|c|c|c|c|c|}
\hline
Model &Texas & Florida & California & Arizona & Georgia & Louisiana & Tennessee & North Carolina & Washington & Pennsylvania\\
\hline
Baseline & 18.00&	48.21&	24.19&	31.35&	29.90&	39.84&	35.88&	19.74&	26.44&	18.70 \\
Baseline + Restaurants & 32.44&	43.84&	21.86&	45.32&	33.46&	29.36&	32.51&	22.91&	23.92&	18.10\\
Baseline + Bars & 44.50	&32.55&	19.89&	26.20&	36.39&	43.51&	38.09&	26.68&	22.75&	24.68\\
\hline
\end{tabular}
}
\end{table*}

Table \ref{tbl:high-eval} shows that regions with a significant causality effect, the RMSE improves on average. CA is an example of such an improvement. 

\begin{table*}[!t]
\centering
\caption{RMSE scores for new cases time series (Baseline), Baseline + Restaurants time series, and Baseline + Bars time series for top-10 regions with the lowest daily new cases in the US.}
\label{tbl:low-eval}
\resizebox{\textwidth}{!}{%
\begin{tabular}{|c|c|c|c|c|c|c|c|c|c|c|}
\hline
Model & Kansas & Hawaii & New Hampshire & Maine & West Virginia & Rhode Island & Connecticut & Montana & Nebraska & Delaware\\
\hline
Baseline & 28.41&	51.49&	12.09&	20.92&	26.18&	5.37&	3.47&	29.58&	5.49&	20.73 \\
Baseline + Restaurants & 25.56&	43.64&	8.10&	14.57&	22.55&	8.88&	3.91&	43.34&	8.22&	20.42\\
Baseline + Bars & 34.43&	49.01&	15.30&	21.96&	24.15&	6.01&	4.68&	43.27&	8.67&	12.81\\
\hline
\end{tabular}
}
\end{table*}

\begin{figure*}[!h]
\centering
\includegraphics[width=\textwidth]{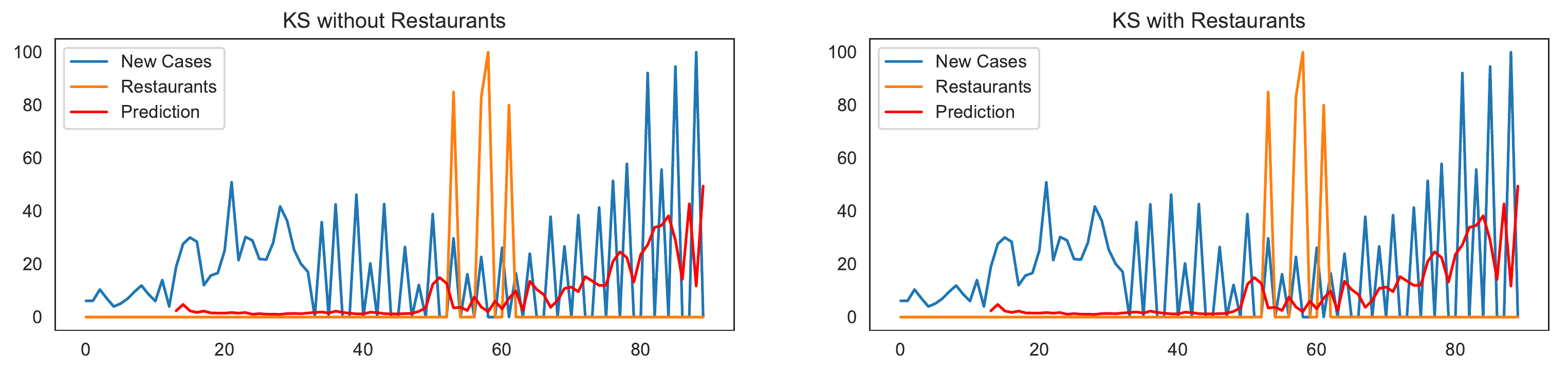}
\caption{Prediction values for daily new cases with/without using restaurant search trends for Kansas (KS).}
\label{fig:ks}
\end{figure*}

\begin{figure*}[!h]
\centering
\includegraphics[width=\textwidth]{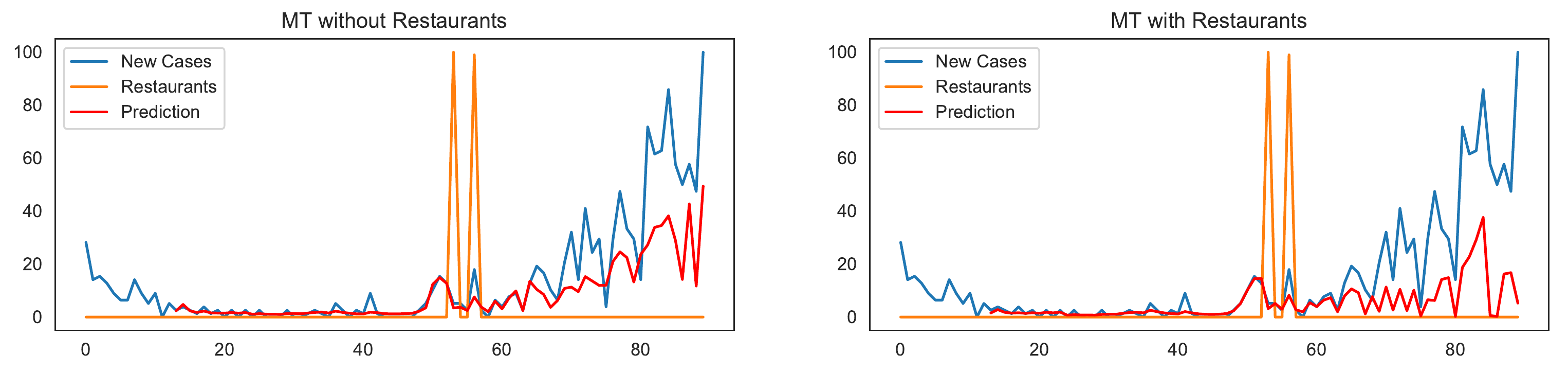}
\caption{Prediction values for daily new cases with/without using restaurant search trends for Montana (MT).}
\label{fig:mt}
\end{figure*}

Table \ref{tbl:low-eval} shows for some states, although there is no causality effect for the restaurants, the value of RMSE improves. On the other hand, for states like Montana (MT), which Granger’s Causality Test shows a significant effect, the RMSE has been increased. By investigating the time series for these two states (Figs. \ref{fig:ks} and \ref{fig:mt}), we can interpret such inconsistencies for two reasons. First, for states such as Kansas (KS), the improved value is because of the fluctuation in the new cases time series, making the prediction unreliable. Second, as Figs. \ref{fig:ks} and \ref{fig:mt} show, the impulses in restaurant searches for KS and MT are point impulses. These unit jumps cannot improve the prediction of the time series, although they appear in causality tests.

\section{DISCUSSION}
\subsection{Principal Results}
To the best of our knowledge, this study is the first analysis that considers the predictive ability of Google search trends, namely restaurants and bars, on daily new cases of COVID-19 in the US. This study uses statistical methods to validate such an effect on daily new cases.

Granger’s causality test shows that in some regions, the effect of restaurants on daily new cases is significant. California is an example of such states. On May 18th, the governor of California announced the easing of criteria for counties to re-open faster than the state, and on May 25th he announced plans for the re-opening of in-store shopping \cite{timelineJohnHopkins}. Consequently, there was an increase in restaurant searches, and the peak of the searches happened on June 7th.  The daily new cases drastically increased within two weeks of the escalation in dine-in restaurant searches. 

Similarly, such a trend in bar searches happened in California (Fig. \ref{fig:comparison}). Regardless of the seasonal effect of time series, which shows a higher number of searches for bars during weekends, the average trend in bar searches increased. However, North Carolina is not influenced by restaurant searches (Table \ref{tbl:high-cause}). The reason is that this state has an increasing average trend regardless of the other time series (Fig. \ref{fig:comparison2}). Therefore the P-value for Granger’s causality is high (.53).

This study suggests that the effect of restaurant and bar searches is higher in the regions with higher daily new cases compared to the regions that have a lower number of positive cases reporting every day. On average, in the regions with a higher number of daily new cases, more significant Granger's casualties and higher values of Pearson correlation support this fact.

We used artificial intelligence models to improve the prediction results of new cases using additional information, namely Google trends. These Google trends for restaurants and bars can be useful depending on the time series structure. Prediction in time series uses the information of previous values (lags) to estimate the current values.

\subsection{Limitations}
There are several limitations to this study. We only used the most popular search queries suggested by Google for each category. People use different search terms to find the information they are looking for. Moreover, we only considered the effect of restaurants and bars on daily cases. The other limitation of our study can be the limited number of samples for each region (88 samples on average). This limitation affects the prediction results to a certain degree.

\section{CONCLUSION}
In conclusion, we investigated the causality effect of search queries related to restaurants and bars on daily new cases in the US regions with high and low daily cases. We showed that for most of the regions with a high number of daily new cases, the effect of search queries on bars and restaurants is higher; hence, they can be used as additional information for prediction tasks.

\addtolength{\textheight}{-12cm}   




\bibliographystyle{IEEEtran}
\bibliography{ref}



\end{document}